\documentclass[conference]{IEEEtran}
\IEEEoverridecommandlockouts
\usepackage{graphicx}
\usepackage[dvipsnames]{xcolor}
\usepackage{amsmath}
\usepackage{amssymb}

\usepackage{mathtools}

\begin{document}
\title{The Case for Asymmetric Systolic Array Floorplanning}

\author{
\IEEEauthorblockN{Christodoulos Peltekis, Dionysios Filippas, Giorgos Dimitrakopoulos} 
\IEEEauthorblockA{Electrical and Computer Engineering\\ 
Democritus University of Thrace, Greece}
\and
\IEEEauthorblockN{Chrysostomos Nicopoulos}
\IEEEauthorblockA{Electrical and Computer Engineering\\  University of Cyprus, Cyprus}}
\maketitle

\begin{abstract}
The widespread proliferation of deep learning applications has triggered the need to accelerate them directly in hardware. General Matrix Multiplication (GEMM) kernels are elemental deep-learning constructs and they inherently map onto Systolic Arrays (SAs). SAs are regular structures that are well-suited for accelerating matrix multiplications. Typical SAs use a pipelined array of Processing Elements (PEs), which communicate with local connections and pre-orchestrated data movements. In this work, we show that the physical layout of SAs should be asymmetric to minimize wirelength and improve energy efficiency. The floorplan of the SA adjusts better to the asymmetric widths of the horizontal and vertical data buses and their switching activity profiles. It is demonstrated that such physically asymmetric SAs reduce interconnect power by 9.1\% when executing state-of-the-art Convolutional Neural Network (CNN) layers, as compared to SAs of the same size but with a square (i.e., symmetric) layout. The savings in interconnect power translate, in turn, to 2.1\% overall power savings.
\end{abstract}

\section{Introduction}
The acceleration of Machine Learning (ML) models~\cite{YOLO, attention} -- both for training and inference -- relies predominantly on matrix multiplications that innately map to systolic arrays~\cite{why-systolic}. Tensor Processing Units (TPUs)~\cite{tpu} and other related architectures~\cite{scalesim, meissa, factored-sa} are notable examples of newly designed SA architectures/derivatives.

Matrix multiplication can be implemented in SAs using integer or Floating-Point (FP) arithmetic~\cite{ten-lessons}. For increased accuracy, the training of deep learning models employs FP arithmetic. To increase energy efficiency, reduced-precision FP formats have been proposed~\cite{bfloat, NIA-fp8} that still offer acceptable accuracy. Nevertheless, when energy efficiency is of principal importance, inference is typically executed using integer arithmetic, after appropriate data quantization of the ML models~\cite{quantization}.

The dynamic power consumption of a SA when computing a matrix multiplication consists of three main components: (a) Input- and weight-loading in the horizontal and vertical directions; (b) Computation power, i.e., the dynamic power consumed for the actual computations of multiplications and additions; and (c) Sum accumulation/unloading that involves the power cost of moving the partial, or final, sums within the columns of the SA. 

This work tackles the \emph{interconnect-related part} of the first and third of the aforementioned components; namely, the consumption attributed to the wires routed in the horizontal and vertical directions across the PEs of the SA and used for (a) the loading of the data and weights and (b) to transfer the partial sums of the reduction operation. Our goal is to select the aspect ratio of the PEs of the SA in a way that minimizes the average dynamic power consumption in said data buses. 

The proposed optimization leverages \emph{the inherent asymmetry} in the width of the horizontal and the vertical data buses used in SAs, as well as the inherent differences in the switching activity profiles of the data objects that pass through these buses. 

The outcome of this optimization is to design \textit{asymmetric} SAs that are more energy-efficient than their typical counterparts, which tend to use `square' processing elements for maximum regularity. Experimental results derived from the physical implementation of various SAs at 28 nm technology demonstrate that the proposed approach yields 9.1\% reduction in the total interconnect power of the design. In turn, this translates into 2.1\% lower overall power consumption within the SA, when executing selected layers of state-of-the-art CNNs.

\section{Matrix Multiplication on Systolic Arrays}

\begin{figure}[thb]
\centering
\includegraphics[width=0.95\columnwidth]{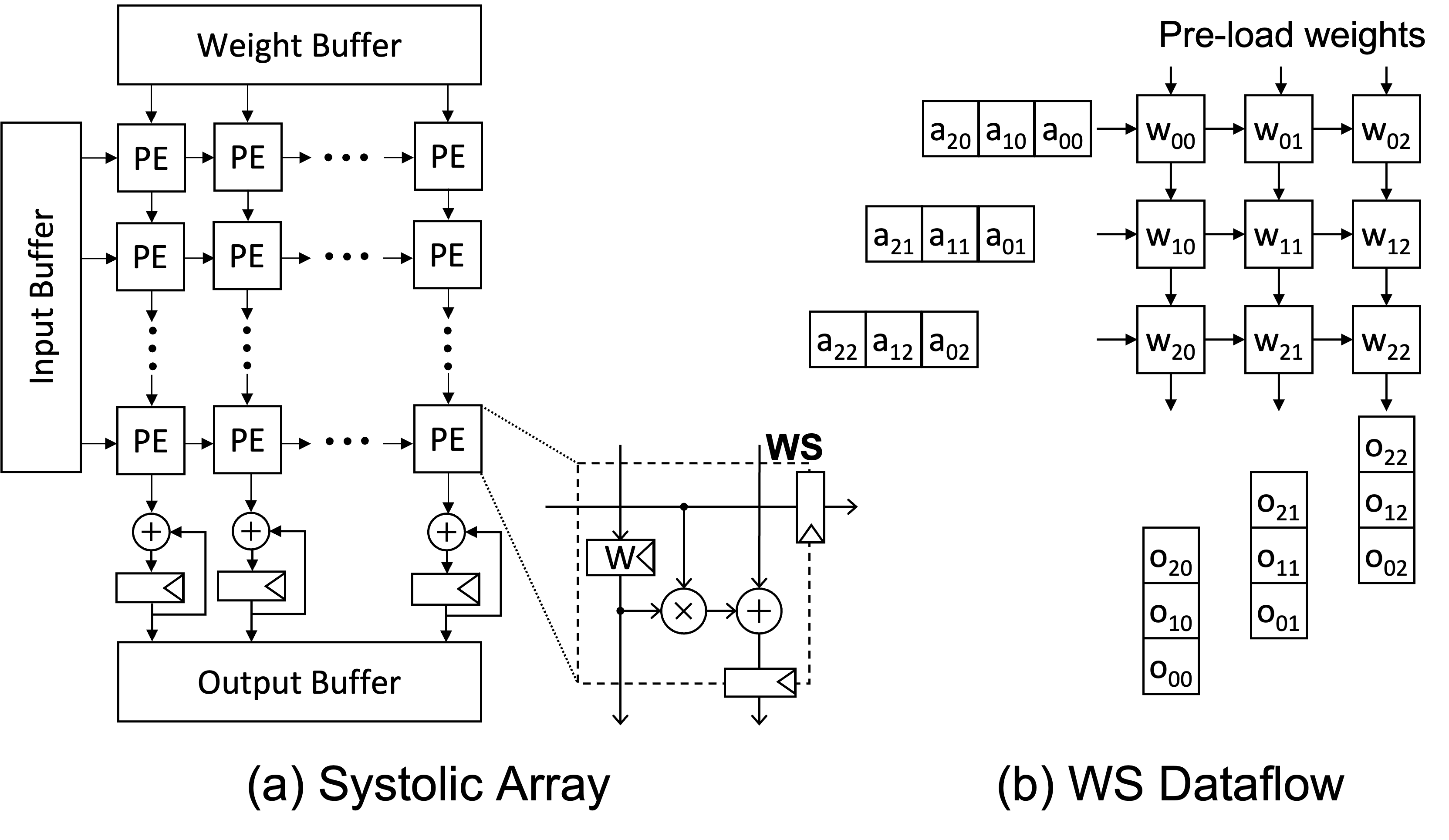}
\caption{The architecture of a generic systolic array, and an abstract overview of the weight-stationary dataflow within the SA.}
\label{f:sa-baseline}
\end{figure}

The most common SA hardware architecture comprises of an array of PEs, as illustrated in Fig.~\ref{f:sa-baseline}(a). Each PE consists of a multiplier, an adder, and various registers to appropriately pipeline the streaming operation. The SA accesses local memory banks situated on the West and North edges of the array, while the output results are collected on the South edge. 

The \textit{dataflow} type utilized by the SA determines the internal micro-architecture of the PEs and how the matrix multiplication, $A\times W$, is executed. For example, in Weight-Stationary (WS) dataflow~\cite{scalesim}, matrix $W$ (containing the weights) is pre-loaded in the SA, while matrix $A$ (containing the input) is transposed and fed into the SA from the West side, as depicted in Fig.~\ref{f:sa-baseline}(b). 
After the top row is filled, it takes multiple cycles to \textit{reduce} the results of all the PEs in the same column; i.e., the result of each PE moves downwards to the next PE in the same column. 

The WS approach is generally preferred over other dataflows, since it exploits the high spatio-temporal reuse of the weights~\cite{tpu}. Nevertheless, to increase flexibility,  configurable SAs can support various dataflow types~\cite{hetero-sa, dataflow-mirroring} and pipelined organizations~\cite{arrayflex}.

Under the WS dataflow, a chain of multiply-add operations is computed in each column of the array. Such consecutive additions inevitably increase the dynamic range of the output result, as compared to the dynamic range of the inputs and weights. For integer-arithmetic implementations, this implies that the adder's output width should be more than double the width of the input (or weights) to retain all accuracy. 

Similarly, in FP implementations, the FP multiply-add units in each PE have a fused/cascaded structure~\cite{galal-fma, aicas}, whereby the product of the multiplication is passed directly to the adder, without intermediate normalization and rounding.
To avoid precision loss, the intermediate results produced at the South output of each PE use double-width precision~\cite{ten-lessons}. For instance, for Bfloat16 inputs, the reduction that occurs in the vertical direction is implemented with FP32 arithmetic. 

Therefore, even if SAs follow an architecturally regular and symmetric structure, the amount of wiring required in the vertical direction -- at least for SAs that support the widely used WS dataflow -- is significantly higher than the wiring of the horizontal direction. 

A similar asymmetry is observed in the switching activity profiles of the inputs that flow horizontally within the rows of the SA and the partial sums that flow vertically in each column of the SA. The input data does not follow a certain statistical pattern and the input values are highly dependent on the selected activation function. The only consistent attribute that is observed for input data is the abundance of zero values generated by the Rectified Linear Unit (ReLU) activation function in each layer. On the contrary, the partial sums that flow in the vertical direction of a SA exhibit more random behavior and possibly higher switching activity, since either positive, or negative, or zero products may need to be accumulated in each step and passed to the next row of the SA.

\section{Optimization of the SA floorplan}

In this work, our goal is to leverage (a) the innate asymmetry in the wiring of the horizontal and vertical directions in a typical SA, and (b) the difference in the switching activity profiles of the data flowing in each direction, in order to design energy-efficient SAs by optimizing their floor-plan design. 

The SA consists of $R$ rows and $C$ columns of PEs. The area of each PE is determined by the area of its constituent components, i.e., the multiplier, the adder (or fused multiply-add units in the case of FP-arithmetic implementations), and the necessary pipeline registers. As shown in Fig.~\ref{f:symetric-assymetric-sa}(a), for a fixed input data width of $B_h$ (in the horizontal direction) and an output data width of $B_v$ (in the vertical direction), we can assume that each PE has a constant area of $A$ $\mu m^2$. Let us also assume that each PE has height $H$ and width $W$. Since the area of each PE is constant, then $H W = A$.

The aspect ratio $W/H$ of the physical layout of each PE can be optimized to minimize the power consumption associated with the wide data buses that run horizontally and vertically inside the SA and connect the PEs in each direction.

\subsection{Minimizing the total wire length}
\label{ss:minimize-wire-length}

The input signals arriving at the West side of each row of the SA are broadcast to all columns of the SA. In practice, those wires are interrupted by pipeline registers at the borders of each PE. However, the inclusion of these registers does not change their overall length. Thus, the bus of $B_{h}$ wires in each row of the SA crosses $C$ columns of PEs of $W \mu m$ width each. The same is true for all rows of the SA. Consequently, the total wire-length of the horizontal input connections of all rows is equal to:
\begin{equation}
WL_{h} = R C ( W B_{h} )
\end{equation}

Similarly, in the vertical direction of each column of the SA, a bus of $B_{v}$ wires runs in the North-to-South direction. This bus connects to the adders of each PE and the corresponding output registers. In each column, this bus spans $R$ PEs with a height of $H \mu m$. Thus, for all columns of the SA, the total wire-length of the vertical connections is equal to:
\begin{equation}
WL_{v} = R C ( H B_{v} ) 
\end{equation}

\begin{figure}
    \centering
    \includegraphics[width=0.95\columnwidth]{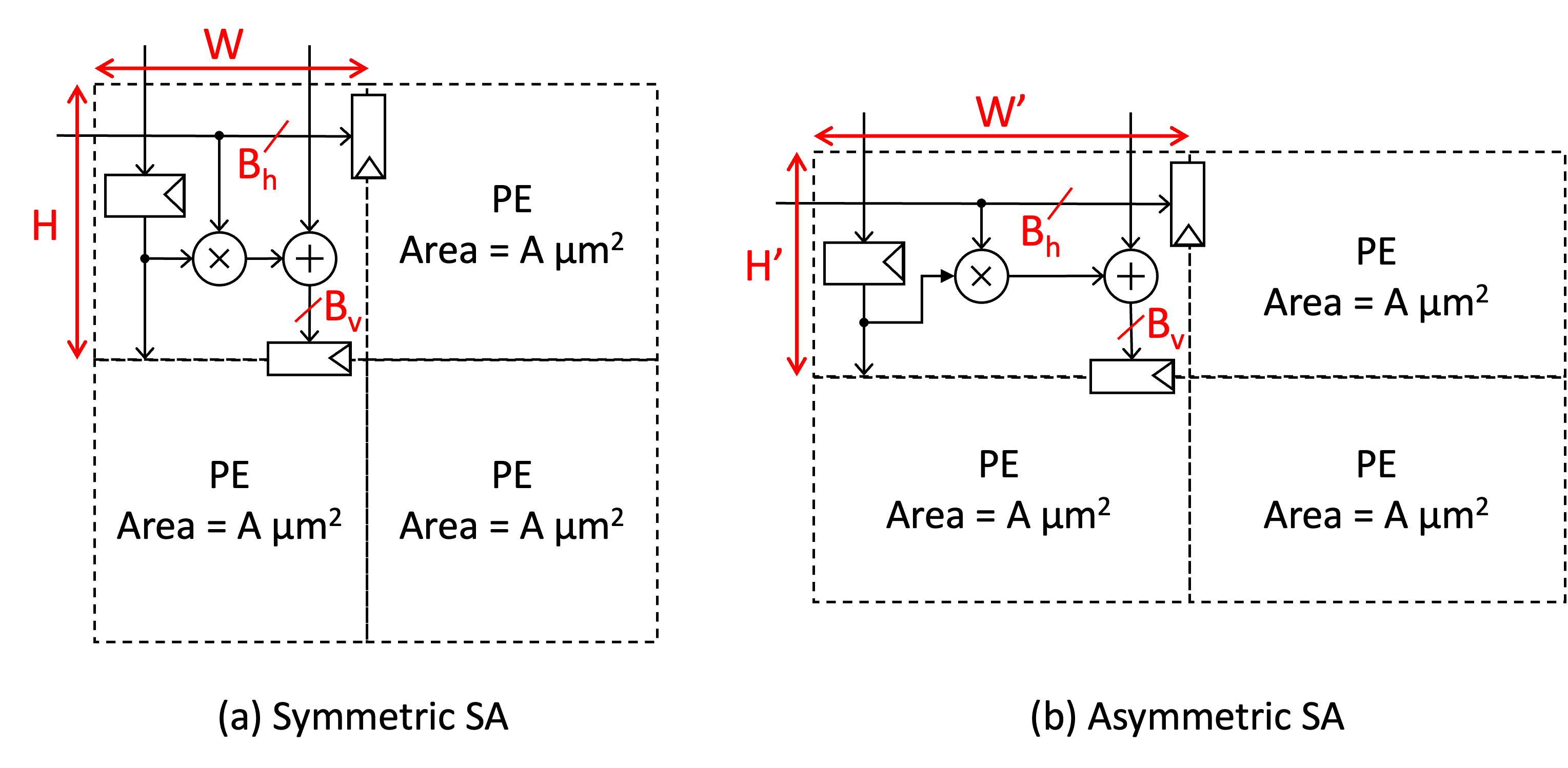}
    \caption{The salient dimensions and bus widths in each PE of (a) a \textit{symmetric}, and (b) an \textit{asymmetric} SA organization.}
    \label{f:symetric-assymetric-sa}
\end{figure}

Adding the wire-lengths of the horizontal and vertical connections/buses, we can calculate the \textit{total} wire-length $WL$:
\begin{equation}
WL = WL_{h} + WL_{v} = RC ( W B_{h} + H B_{v} )
\label{e:wl}
\end{equation}
Substituting  $A/H$ for $W$ in~(\ref{e:wl}),
\begin{equation}
WL = RC \left ( \frac{A B_{h}}{H} + H B_{v} \right ) 
\label{e:wl-new}
\end{equation}    

To identify the optimal aspect ratio that minimizes $WL$ for a PE of constant area $A$, one must take the derivative of $WL$ with respect to $H$
and set it equal to zero. This leads to:
\begin{equation}
\frac{A B_{h}}{ H^2} = B_{v} 
\xrightarrow[]{A = H W}
\frac{W H}{H^2} = \frac{B_v}{B_h}  
\to \frac{W}{H} = \frac{B_v}{B_h}   
\end{equation}
Hence, the aspect ratio $W/H$ that minimizes the total wire-length follows the ratio of the width of the vertical and horizontal buses. Since the vertical connections that carry partial sums are -- by construction -- wider than the horizontal buses that transfer input data, the presented analysis indicates that the PEs should \textit{not be square}. Instead, they should adopt a \textit{rectangular shape} with smaller height than width, as depicted in Fig.~\ref{f:symetric-assymetric-sa}(b), where $H^{\prime} < W^{\prime}$.

In summary, this analysis demonstrates that the aspect ratio of each PE's floorplan should be \textit{asymmetric}, if one wishes to optimize the energy efficiency. This result holds for \textit{all} SAs, irrespective of their size, i.e., the number of rows and columns.

\subsection{Optimal floorplan aspect ratio}

Instead of dealing only with the minimization of the wire length, the above analysis could be enhanced to also include the average switching activity profiles of (a) the input data flowing in the horizontal direction and (b) the partial sums flowing in the vertical direction. In this way, the optimization of the SA floorplan could also target the dynamic power consumption within the interconnects that traverse the PEs in both directions.

To include the average switching per bit in the analysis, it suffices to scale the bit widths of the horizontal and vertical busses, $B_h$ and $B_v$, respectively, with the average switching activity observed in each direction (i.e., $a_h$ and $a_v$) when executing representative ML matrix multiplication workloads. In this case, $B_h$ is scaled by $a_h$ and $B_v$ by $a_v$. After applying such scaling factors, we arrive at the following optimal aspect ratio for each PE:
\begin{equation}
\frac{W}{H} = \frac{B_{v} a_{v}}{B_{h} a_{h}}
\label{e:final}
\end{equation}

Note that this result does not change the main outcome of the analysis presented in Section~\ref{ss:minimize-wire-length} above. Each PE should still have a smaller height than width, since $a_h$ is expected to be smaller, or at least equal to, $a_v$. Therefore, the fact that $B_v > B_h$ still determines the optimal aspect ratio.

\section{Experimental Evaluation}

To highlight the benefits of asymmetric SA floorplanning, we designed two 32$\times$32 SA-based accelerators in SystemVerilog Register-Transfer Level (RTL). The SAs were implemented with the Cadence digital implementation flow using a 28 nm standard-cell library. The first accelerator has a fully symmetric layout, where each PE is placed as a square, while the second SA adopts the proposed asymmetric floorplan with rectangular PEs.

Both SA designs operate at 1 GHz with 16-bit integer quantized inputs and weights, and they both execute single-batch inference on the ResNet50~\cite{resnet} CNN layers, which consist of matrix multiplications of different sizes. The additions in each column of the SAs are performed at a width of 37 bits. This particular output bit-width is required to accommodate the dynamic range when adding 32 products of 32 bits each. Therefore, for the selected configuration, $B_h$=16 and $B_v$=37.

The switching activity information was collected after feeding the CNN model with sample images from ImageNet~\cite{imagenet}. The average switching activity observed after executing all layers of ResNet50~\cite{resnet} is $a_h$=0.22 and $a_v$=0.36. It is important to note that the horizontal switching activity is an average of the input observed in all layers of ResNet50. This means that layers with denser inputs have higher switching activity and layers with sparser input (i.e., more zero values observed due to the ReLU activation function) have lower switching activity. The increased average switching activity observed in the vertical direction (output of the adder of each PE) is mostly due to the signed arithmetic that causes many bits to change value when moving from positive to negative numbers. On the contrary, the inputs in the horizontal direction are, by construction, positive integers.

These switching activities are merely used as indicative examples. For a real design, one needs to take into account the switching profiles of many applications, in order to arrive at a solution that is efficient in various different application scenarios.

\begin{figure}[h!]
    \centering
    \includegraphics[width=0.55\columnwidth]{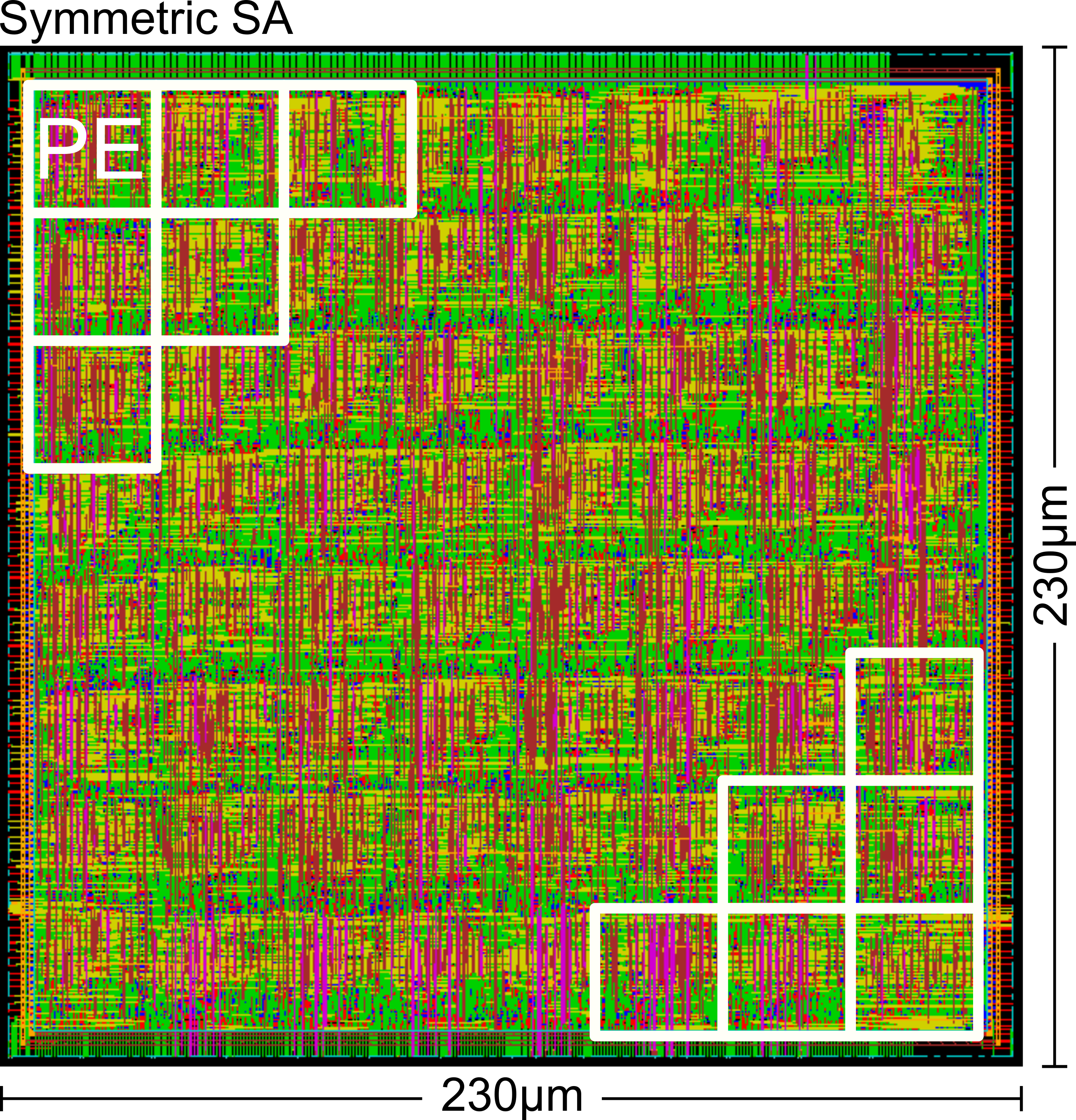} \\
    {\small (a)} \\
    \includegraphics[width=0.85\columnwidth]{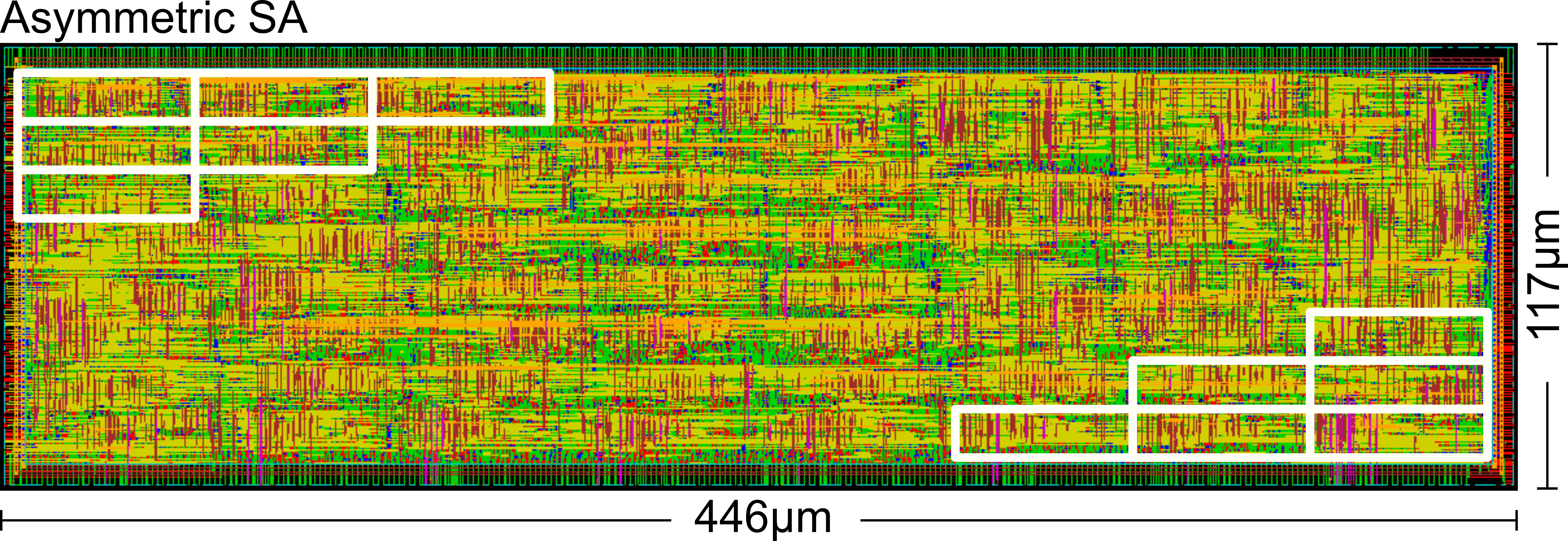} \\
    {\small (b)}
    \caption{Physical layouts of the 8$\times$8 (a) symmetric SA, and (b) asymmetric SA. Both designs were implemented using a 28 nm standard-cell library and operate at a clock frequency of 1 GHz.}
    \label{f:layout}
\end{figure}

Based on the width of the horizontal and vertical data buses, and the observed switching activities (that represent the average behavior of many CNN models), we selected an aspect ratio of $W/H = 3.8$ for the proposed asymmetric SA design. Fig.~\ref{f:layout} shows the physical layouts of a conventional SA with square PEs and the proposed asymmetric SA design with rectangular PEs, assuming arrays with 8$\times$8 PEs.

\begin{table}
    \centering
    \renewcommand{\arraystretch}{1.1}
    \caption{Selected convolutional layers of ResNet50~\cite{resnet} and their respective attributes. Parameter K is the kernel size, H is the output height, W is the output width, C is the number of input channels, and M is the number of output channels.}
    \label{t:layers}
    \begin{tabular}{|c|c|}
        \hline
         Name & Attributes  \\\hline\hline
         L1   &  K=1, H=56, W=56, C=256,  M=64  \\\hline
         L2   &  K=3, H=28, W=28, C=128,  M=128 \\\hline
         L3   &  K=1, H=28, W=28, C=128,  M=512 \\\hline
         L4   &  K=1, H=14, W=14, C=512,  M=256 \\\hline
         L5   &  K=1, H=14, W=14, C=1024, M=256 \\\hline
         L6   &  K=3, H=14, W=14, C=256,  M=256 \\\hline
    \end{tabular}
\end{table}

\begin{figure}[h!]
    \centering
    \includegraphics[width=0.7\columnwidth]{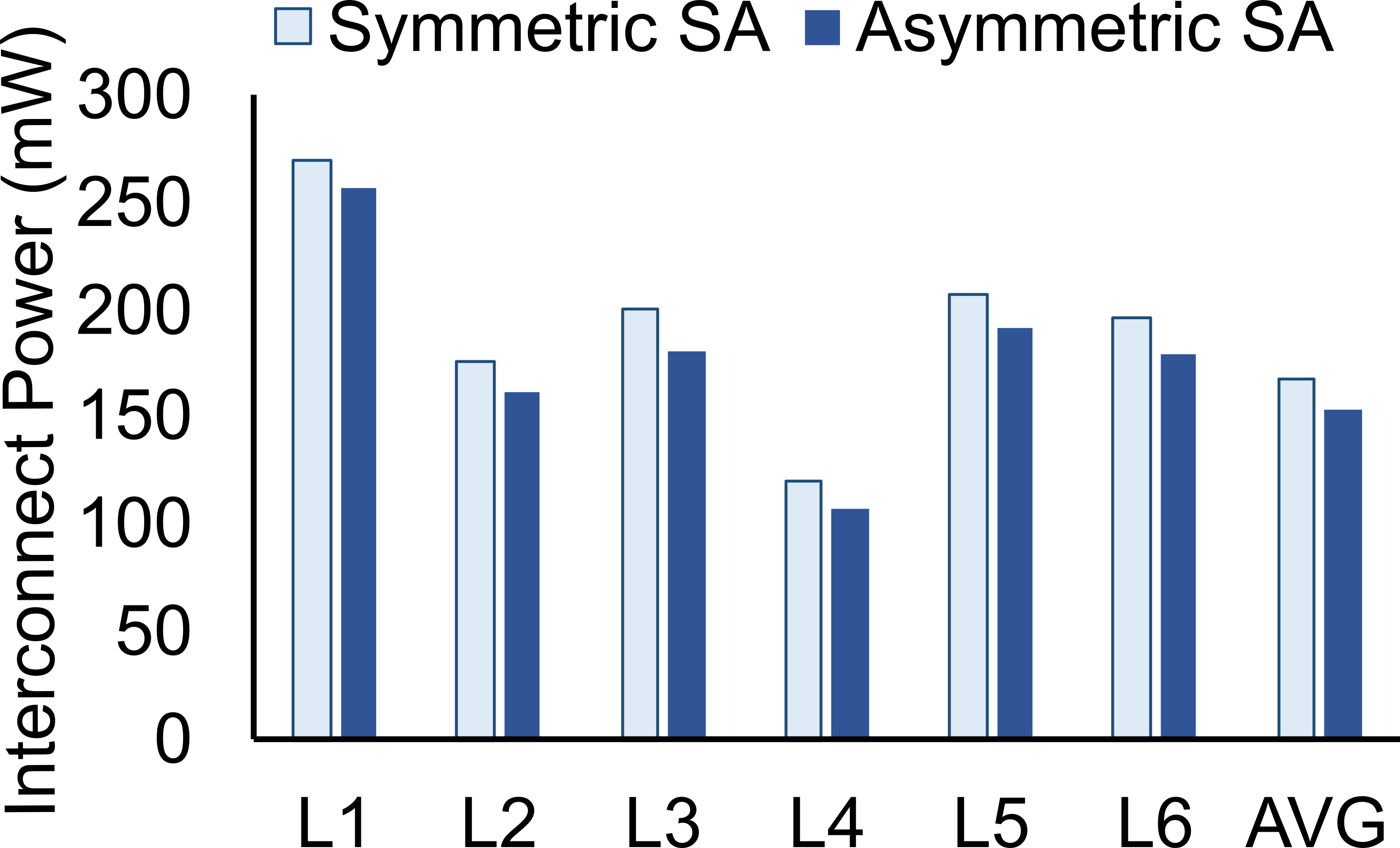}
    \caption{Interconnect power consumption for selected layers of ResNet50~\cite{resnet}.}
    \label{f:intercon-power}
\end{figure}

\begin{figure}[h!]
    \centering
    \includegraphics[width=0.7\columnwidth]{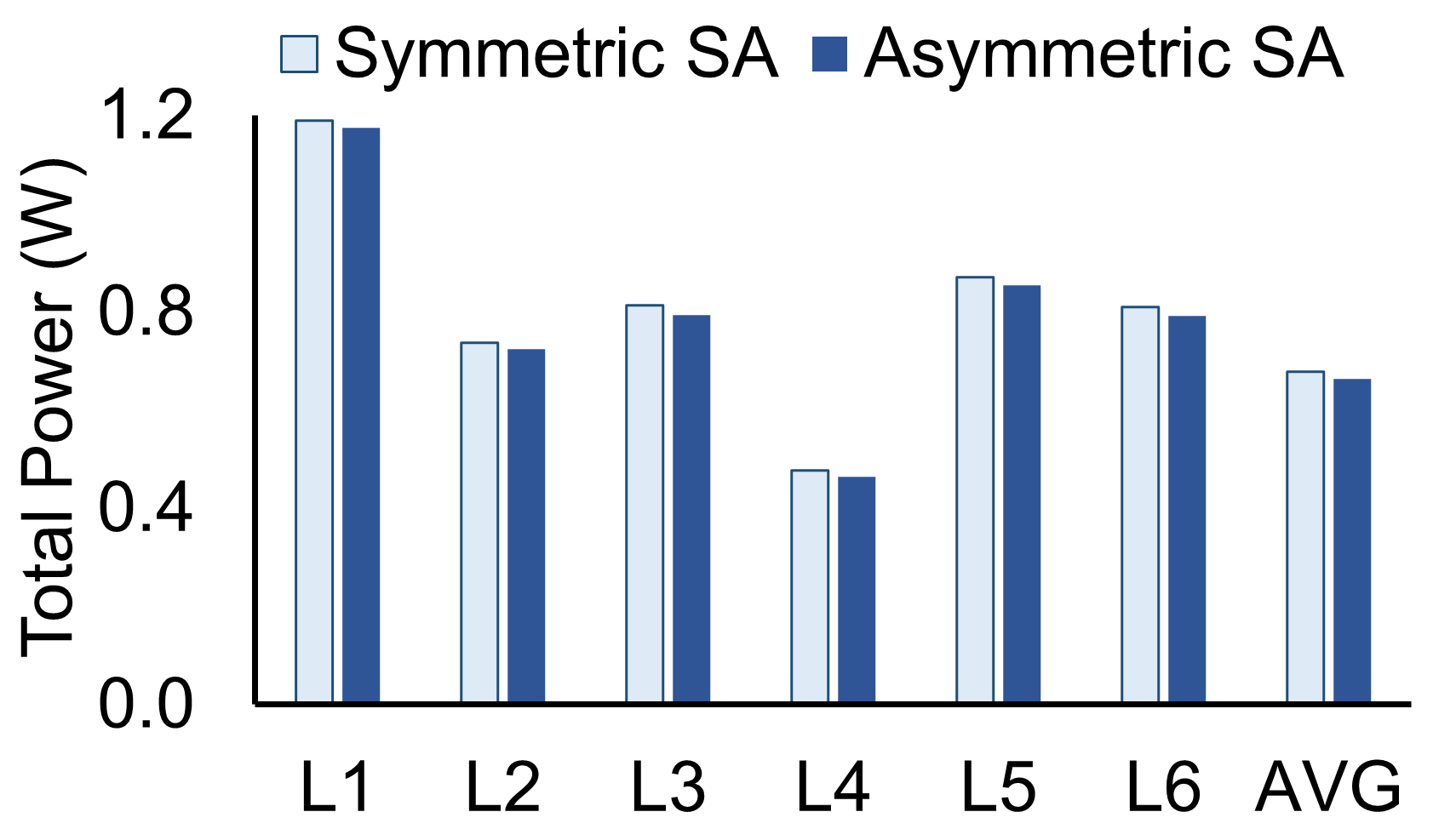}
    \caption{Total power consumption for selected layers of ResNet50~\cite{resnet}.}
    \label{f:total-power}
\end{figure}

Fig.~\ref{f:intercon-power} depicts the power consumed on the interconnects of the symmetric and asymmetric SA configurations, for 6 selected convolutional layers of ResNet50~\cite{resnet}. The average per-layer power consumption for ResNet50 is also shown in the figure. The corresponding dimensions of the 6 selected layers are summarized in Table~\ref{t:layers}. As shown in Fig.~\ref{f:intercon-power}, the proposed asymmetric layout reduces the total interconnect power consumption by 9.1\%, as compared to the symmetric layout. This interconnect power reduction translates into a \textit{total} average power reduction of 2.1\%, as illustrated in Fig.~\ref{f:total-power}.

Even if the overall power savings are small, they are reaped without \textit{any} performance trade-off whatsoever. The proposed approach simply requires a customization of the design's floorplan during its physical synthesis flow.

\section{Conclusions}

Systolic arrays embrace regularity both in their physical structure and in their cycle-by-cycle operation. The only asymmetry that they exhibit is in the required amount of horizontal (input) wiring, relative to the vertical (output) wiring. In this work, we leverage this physical asymmetry and the difference in the switching activity profiles of the data flowing in the horizontal and vertical directions to optimize the interconnect-related dynamic power consumption of the SA. The proposed optimization targets the physical floorplanning of the PEs of the SA, and it allows for improvements in the total dynamic power of the design. The proposed approach is complementary to other data-driven low-power techniques proposed for SAs~\cite{mocast-lp-sa}, thereby further increasing the energy efficiency of the design. These optimizations could lead to more power-efficient neural network accelerators, beneficial for applications with power limitations like ambient assisted living systems developed in domestic environments.

\section*{Acknowledgements}
This work is supported by the project "Study, Design, Development and Implementation of a Holistic System for Upgrading the Quality of Life and Activity of the Elderly" (MIS 5047294) implemented under the Action "Support for Regional Excellence", funded by the Operational Programme "Competitiveness, Entrepreneurship and Innovation" (NSRF 2014-2020), which is co-financed by Greece and the EU (European Regional Development Fund).

\bibliographystyle{IEEEtran}
\bibliography{refs}

\end{document}